\documentclass[11pt, a4paper]{article}

\usepackage[utf8]{inputenc}
\usepackage[T1]{fontenc}

\usepackage{xcolor}
\usepackage{graphicx}
\usepackage{hyperref}
\usepackage{url}
\usepackage{booktabs}
\usepackage{amsmath, amssymb}
\usepackage{bm}
\usepackage{natbib}
\usepackage{indentfirst}
\usepackage{multicol}
\usepackage{float}
\usepackage{times}

\usepackage[bottom=2cm, left=3cm, right=2cm, top=3cm, headheight=14pt]{geometry}

\usepackage{fancyhdr}
\fancypagestyle{firstpagestyle}
{
    
    \fancyhf{}
    \rfoot[R]{\footnotesize Page \thepage}
}

\newcommand{\thetitle}{Dynamic Clustering of Time Series Data}

\begin{document}

\thispagestyle{firstpagestyle}

\begin{center}
    \rule{15.5cm}{2pt}
    \vspace{10pt}

    \textsc{\textbf{\LARGE \thetitle}}

    \rule{15.5cm}{2pt}
\end{center}

\begin{multicols}{2}
    \begin{center}
        Victhor S. Sartório \\
        Instituto de Matemática \\
        Universidade Federal do Rio de Janeiro \\
        Rio de Janeiro, Brazil \\
        \texttt{victhor@dme.ufrj.br}
    \end{center}

    \vfill{\null}
    \columnbreak 

    \begin{center}
        Thais C. O. Fonseca \\
        Departamento de Métodos Estatísticos \\
        Universidade Federal do Rio de Janeiro \\
        Rio de Janeiro, Brazil \\
        \texttt{thais@im.ufrj.br}
    \end{center}
\end{multicols}

\begin{abstract}
    We propose a new method for clustering multivariate time-series data based on Dynamic Linear Models. Whereas usual time-series clustering methods obtain static membership parameters, our proposal allows each time-series to dynamically change their cluster memberships over time. In this context, a mixture model is assumed for the time series and a flexible Dirichlet evolution for mixture weights allows for smooth membership changes over time. Posterior estimates and predictions can be obtained through Gibbs sampling, but a more efficient method for obtaining point estimates is presented, based on Stochastic Expectation-Maximization and Gradient Descent. Finally, two applications illustrate the usefulness of our proposed model to model both univariate and multivariate time-series: World Bank indicators for renewable energy consumption of EU nations and the famous Gapminder dataset containing life-expectancy and GDP per capita for various countries.
\end{abstract}

\textbf{\textit{Keywords}} \hspace{10pt} Time-series $\cdot$ Clusterization $\cdot$ Mixture Models $\cdot$ Dynamic Models


\baselineskip18pt

\section{Introduction}

Clusterization is the process grouping data into different sets, called \textit{clusters}, that are heterogeneous from each other, but such that members of the same cluster are relatively similar among themselves. This approach is used for identifying structure in unlabeled datasets and has become increasingly popular for both exploratory analysis and automatic segmentation of data points.

Historically, most clustering techniques concern themselves with static data. With the rise of routine collection of data, however, attention has been increasingly turned to clustering time-varying observations. Although the number of developments in this field for the past few decades can seem overwhelming,~\cite{liao2005clustering} and~\cite{aghabozorgi2015time} serve as good summaries for the most significant publications.

Of particular interest for time-series clustering are Hidden Markov Models (HMMs). They make up a general framework of probabilistic time-series models that rely on abstracting away temporal dependencies into unobserved states whose evolution follow the Markov property. The idea of clustering time-series with HMMs was initially presented by~\cite{juang1985probabilistic} and made significant impact in fields such as speech recognition \citep{juang1991hidden}, protein modelling \citep{krogh1994hidden}, and genetics \citep{schliep2003using}, among others.

\begin{figure}[H]
    \centering

    \includegraphics[width=0.5\linewidth]{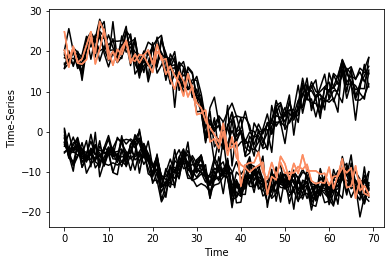}
    \caption{Illustrative example of time-series spread across two clusters where two of the time-series, colored in orange, change their cluster membership halfway through the time window.}\label{fig:s1_example}
\end{figure}

When clustering time-series data that span long windows of time, however, it may be inappropriate to assume that every single observation will follow the behavior of the same cluster for the whole period. Consider Figure~\ref{fig:s1_example} for an illustrative example of what this behavior would look like.

Evidently, one could argue that such behaviour is that of a new, third, cluster. However, as more observational units start presenting this behavior at different time-points, one would require a big number of clusters with very few members in each of them to capture these changes, which, in a way, goes against the very purpose of clustering.

For a model to succesfully acommodate such behavior, it would need to account for the possibility of time-series changing their cluster memberships through time, and little work has been done in this regard in existing literature.

There have been two particularly relevant publications with this approach. First,~\cite{punzo2016clustering} proposed a Markov chain dictating the transition probabilities between clusters across each longitudinal observation. There are two downsides to this approach: first, the cluster membership changes can be too abrupt; and second, assuming that all observations have the same transition probabilities can be too strict.

On the other hand,~\cite{maruotti2016time} did not use a hidden Markov structure and proposed a model which takes the idea of k-means clustering \citep{macqueen1967some} and adapts it so that each cluster's centroid is ruled by a Vector Autoregressive model. Although a considerably more ideal approach for longer time-series data, the cluster membership itself does not incorporate time-dependence, which can allow for some clear missclassifications of outlier observations as is shown in the Appendix~\ref{sup:app_to_art_dat}.

In this paper, we present clusterization of time-series based on Dynamic Linear Models \citep{west1996bayesian} which are a particular class of HMMs. Then, we further extend this initial model to take into account dynamic mixture coefficients and assume their evolution is ruled by an Evolutional Dirichlet Process \citep{fonseca2017dynamic}, allowing cluster memberships to shift smoothly through time. Two estimation approaches are considered: Markov Chain Monte Carlo inference, which has a considerable higher computational cost but allows for any probabilistic inference to be made for the posterior distributions, and pontual estimation methods based on Stochastic Expectation Maximization and Gradient Descent, providing results much more quickly, but restricting analysis to simple point estimates for model parameters. An accompanying Python library named \emph{dynmix}\footnote{Acessible from the PyPI repositories.} was developed.

In Section 2, we present our model for simple mixtures of DLMs, where the pontual estimation procedure may prove useful in regular HMM applications, and in Section 3 we present its extension for dynamic mixtures of DLMs,. In Section 4 we showcase an application for univariate series regarding renewable energy consumption of EU member states, and a multivariate application considering Life Expectancy and GDP per capta data for European and African countries. In Section 5 we present some final remarks.

\section{Mixtures of Dynamic Linear Models}

Consider that one has observed $n$ different $m$-variate time series across the same $T$ time instants, denoted each by $\bm{y}_i = (\bm{y}_{i1}, \dots, \bm{y}_{iT})$, $i = 1, \dots, n$. Let's then assume that there are $k$ different clusters, indexed by $j = 1,\dots, k$, each with a behavior determined by $p_j$-variate time-varying states $\bm{\theta}_j = (\bm{\theta}_{j1}, \dots, \bm{\theta}_{jT})$ and cluster specific parameters $\bm{\Phi}_j$. Finally, assume that all the dependence in the observed data, across time \textit{and} across different observations, is captured by these states.

Thus, we can consider the following observational model for each of the $n$ conditionally independent observations
\begin{equation}
    \label{eq:s2_obs_general}
    f(\bm{y}_{i} | \bm{\eta}_i, \bm{\theta}_1, \dots, \bm{\theta}_k, \bm{\Phi}_1, \dots, \bm{\Phi}_k) = \sum_{j=1}^k \eta_{ij} \prod_{t=1}^T f_j(\bm{y}_{it} | t, \bm{\theta}_{jt}, \bm{\Phi}_j)
\end{equation}
where $\bm{\eta}_{i} = (\eta_{i1}, \dots, \eta_{ik})$ is the cluster membership vector (or mixture weights) for the $i$-th observation and $f_j$ is the appropriate observational probability density function (p.d.f.) for the $j$-th cluster.

The assumptions made, and consequently equation (\ref{eq:s2_obs_general}), specify an observational model for a general time-series clusterization approach based on Hidden Markov Models, which~\cite{juang1985probabilistic} first proposed.

We now proceed with our choice of Dynamic Linear Models, or DLMs for short, to serve as the specific framework of Hidden Markov Models to be used for modelling the states of each of the clusters. DLMs provide powerful and convenient set tools for flexibly modeling various behaviours.

In particular, a DLM for the states $\bm{\theta}_j$ with observations $\bm{x}_{1}, \dots, \bm{x}_T$ is modelled through the equations
\begin{alignat*}{2}
    \bm{x}_{t} &= \bm{F}_{jt} \bm{\theta}_{jt} + \bm{\nu}_{jt} &
    ~~~~~~ \bm{\nu}_{jt} \sim N(\bm{0}_m, \bm{V}_{jt}) \\
    \bm{\theta}_{jt} &= \bm{G}_{jt} \bm{\theta}_{j,t-1} + \bm{\omega}_t^{(j)} &
    ~~~~~~ \bm{\omega}_t^{(j)} \sim N(\bm{0}_{p_j}, \bm{W}_{jt})
\end{alignat*}
where $\bm{F}_{j}$ is called the \textit{observational matrix}, $\bm{G}_{j}$ is called the \textit{evolutional matrix}, $\bm{V}_{jt}$ is the observational error covariance matrix and $\bm{W}_{jt}$ is the evolutional error covariance matrix.

For the remainder of the paper, to lighten the notation, the time index of the observational and evolutional matrices will be dropped whenever they are not explicitly assumed to be time-varying.

Different specifications of the pair $(\bm{F}_j, \bm{G}_j)$ can be made to take into account various interesting properties from the data, such as different forms of trend, seasonality or dependence on time-varying covariates. See~\cite{west1996bayesian} for a thorough overview of Dynamic Linear Models.

We will also consider indirect modelling of the $\bm{W}_t$ through discount factor modelling \citep{harrison1987practical}. The use of discount factors turn out to be essential for the success of this proposal as this will allow the maximum a posteriori procedure to be well defined for Dynamic Linear Models. This is better detailed in Section~\ref{sec:static_mixture_em}. In summary, the idea of discount modelling is to impose a (known) rate of loss of information regarding the states at each time-step which will lead to a $\bm{W}_t$ that is function of all the other parameters and observations. This strategy for modelling the evolutional variance is also commonly used in practice; see~\cite{lamon1998forecasting} and~\cite{lopes2007factor} for interesting examples.


Now, with the added assumption that the states for each cluster can be modelled through a DLM, we have that the only cluster specific parameter is $\bm{\Psi}_j = \{\bm{V}_j\}$ and the observational equation from (\ref{eq:s2_obs_general}) becomes
\begin{equation}
    \label{eq:s2_obs_weighted}
    f(\bm{y}_{i} | \bm{\eta}_i, \bm{\theta}_1, \dots, \bm{\theta}_k, \bm{V}_1, \dots, \bm{V}_k) = \sum_{j=1}^k \eta_{ij} \prod_{t=1}^T N_m(\bm{y}_{it} ; \bm{F}_j \bm{\theta}_{jt}, \bm{V}_j)
\end{equation}
where $N_p(\cdot; \bm{m}, \bm{C})$ is the probability density function of the $p$-variate normal distribution with mean $\bm{m}$ and covariance matrix $\bm{C}$.

Although equation (\ref{eq:s2_obs_weighted}) fully reflects the observational model of interest, when estimation is concerned it is useful to perform a commonly used model augmentation \citep{tanner1987calculation}. In particular, we introduce a dummy variable $Z_i$ for each observation such that $\Pr(Z_i = j | \bm{\eta}_i) = \eta_{ij}$. By doing this, the observational equation simplifies to
\begin{equation}
    \label{eq:s2_obs_dummy}
    f(\bm{y}_{i} | Z_i, \bm{\theta}_1, \dots, \bm{\theta}_k, \bm{V}_1, \dots, \bm{V}_k) = \prod_{t=1}^T N_m(\bm{y}_{it} ; \bm{F}_{Z_i} \bm{\theta}_{Z_i t}, \bm{V}_{Z_i}) \text{.}
\end{equation}

Although the models presented here will assume that the number of clusters $k$ is known, specially because of the focus on imposing meaningful structures for each cluster DLM, works such as~\cite{richardson1997bayesian} and~\cite{nobile2007bayesian} pave the way for future extensions which could also estimate $k$. Another approach is to perform model comparison for different values of $k$, using metrics such as the modified BIC proposed by~\cite{fraley2007bayesian}.

Now that we have specified a model, we'll go into estimation of the model parameters, for which we present two options: a fully Bayesian approach in which we obtain samples from the posterior distribution by using Markov Chain Monte Carlo (MCMC) methods, and a maximum a posteriori approach which will provide point estimates in a very efficient manner by using the Expectation Maximization (EM) algorithm.

\subsection{Gibbs sampling}

An MCMC estimation approach for this model is made very easy thanks to its dependence structure, which leads to various conditional independences, and also thanks to the analytical tractability of Dynamic Linear Models.

The Gibbs algorithm \citep{geman1987stochastic} involves separating all random variables of interest into different blocks and sampling from the full conditional distribution of each block, one at a time, at each iteration. The resulting stochastic process is guaranteed to converge to the joint posterior distribution of the random variables of interest.

Firstly, assuming $\bm{\eta}_i \sim Dirichlet(\bm{c}_0)$ as the prior, the conditional distribution of each $\bm{\eta}_i$ given all other information can be easily found to be $\bm{\eta}_i | \cdot \sim Dirichlet(\bm{c}_0 + \bm{e}_{Z_i})$ where $\bm{c}_0$ comes from the prior and $\bm{e}_j$ is the $j$-th canonical vector of $\mathbb{R}^k$.

Then, we obtain the conditional distribution of each $Z_i$, which is given by
\begin{equation}
    \label{eq:s2_dummybayes}
    P(Z_i = j | \bm{\eta}_i, \bm{\theta}_1, \dots, \bm{\theta}_k, \bm{V}_1, \dots, \bm{V}_k, \bm{y}_i) = \frac{\eta_{ij} \prod_{t=1}^T N_m(\bm{y}_{it} ; \bm{F}_{j} \bm{\theta}_{jt}, \bm{V}_{j})}{\sum_{l=1}^k \eta_{il} \prod_{t=1}^T N_m(\bm{y}_{it} ; \bm{F}_{l} \bm{\theta}_{lt}, \bm{V}_{l})}
\end{equation}
which, although intuitive and simple to obtain algebrically, may lead to serious computational problems arrising from underflow, specially for higher values of $T$.

A strategy for dealing with this problem is, for each $i$, evaluating all $N_m(\bm{y}_{it}; \bm{F}_j \bm{\theta}_{jt}, \bm{V}_j)$, for all $j$ and $t$, and obtaining the mean order of magnitude
\begin{equation*}
    \bar{O}_{ij} = \frac{1}{T} \sum_{t=1}^T \log_{10} N_m(\bm{y}_{it}; \bm{F}_j \bm{\theta}_{jt}, \bm{V}_j)
\end{equation*}
for each of the clusters. Then, we define a new quantity as $\bar{O}_i = \max_{j = 1, \dots, k} \bar{O}_j$.

Given this value, computing the expression
\begin{equation}
    \label{eq:s2_computational_formula}
    P(Z_i = j | \bm{\eta}_i, \bm{\theta}_1, \dots, \bm{\theta}_k, \bm{V}_1, \dots, \bm{V}_k, \bm{y}_i) = \frac{\eta_{ij} \prod_{t=1}^T 10^{-\bar{O}_i} N_m(\bm{y}_{it} ; \bm{F}_{j} \bm{\theta}_{jt}, \bm{V}_{j})}{\sum_{l=1}^k \eta_{il} \prod_{t=1}^T 10^{-\bar{O}_i} N_m(\bm{y}_{it} ; \bm{F}_{l} \bm{\theta}_{lt}, \bm{V}_{l})}\text{,}
\end{equation}
obviously should lead to the same mathematical result but avoids underflow by keeping the terms of the products of the most relevant clusters around 1.

Conditional on $Z_1, \dots, Z_n$, let $\mathbb{I}_j = \{i : Z_i = j\}$ be the set of observational indexes belonging to the $j$-th cluster and $n_j$ the number of elements of this set.

For each of the $m$ observational dimensions there will generally be state components associated with their behaviour, e.g. their level. By using discount factors, systematic correlations between these dimensions will be captured through the states, and it becomes reasonable to assume that the independent observational discrepancies from the mean at each time-point will be independent from each other. As such, we will assume throughout the paper that $\bm{V}_j$ is diagonal, such that $\bm{V}_j = diag(\phi_{j1}^{-1}, \dots, \phi_{jm}^{-1})$. Thus, we have that the conditional distribution for each of these elements is
\begin{equation*}
    \phi_{jl} | \cdot \sim Gamma\left(\frac{n_j T}{2}, \frac{\sum_{i \in \mathbb{I}_j} \sum_{t=1}^T {\left({(\bm{F}_j \bm{\theta}_{tj})}_l - y_{itl} \right)}^2}{2}\right)
\end{equation*}
where ${(\bm{F}_j \bm{\theta}_{tj})}_l$ denotes the $l$-th element of the resulting vector from the product $\bm{F}_j \bm{\theta}_{tj}$. Note that if one desires to consider observational correlations, the conditional distribution for $\bm{V}_j$ still holds analytical form as an Inverse Wishart distribution.

As for the states, for each cluster $j$ we have a specification which involves an $m$-variate observation and a $p_j$-variate state dimension. However, given $\mathbb{I}_j$, we know we have $n_j$ replicates of that $m$-variate observation.

This is equivalent to considering an $n_j m$-variate observation vector $\bm{Y}^{(j)}_t$ at each time instant, where $\bm{Y}^{(j)}_t$ is a vertical concatenation of the $n_j$ vectors $\bm{y}_i$, $i \in \mathbb{I}_j$. Then, we create a new observational matrix which is a vertical tiling of $n_j$ replicates of the original matrix $\bm{F}_j$ and a new observational error covariance matrix which is given by $\bm{I}_{n_j} \otimes \bm{V}_j$ where $\bm{I}_n$ is the $n \times n$ identity matrix and $\otimes$ denotes the Kronecker product. Then, one can proceed to perform usual forward filtering and backwards sampling for Dynamic Linear Models using these quantities for each cluster $l$, thus, obtaining samples from the conditional distribution of $\bm{\theta}_j$.

\subsection{Expectation Maximization}\label{sec:static_mixture_em}

Obtaining point estimates for finite mixture models is a very common application of the Expectation-Maximization algorithm \citep{dempster1977maximum} due to its convergence properties and computational efficiency.

Considering both $\bm{y}_1, \dots, \bm{y}_n$ and $Z_1, \dots, Z_n$ to be observations, our log-likelihood function is
\begin{align}
    \label{eq:s2_loglik}
\begin{split}
    l(\cdot | \bm{y}_1, \dots, \bm{y}_n, Z_1, \dots, Z_n) = & \sum_{i=1}^n \sum_{j=1}^k \delta(Z_i - j) \left(\ln \eta_{ij} + \ln N_m(\bm{y}_{it} ; \bm{F}_{j} \bm{\theta}_{jt}, \bm{V}_{j})\right) + \\ & \sum_{j=1}^n \sum_{t=2}^T \ln N_{p_j}(\bm{\theta}_{jt}; \bm{G} \bm{\theta}_{j,t-1}, \bm{W}_{jt})
\end{split}
\end{align}
where $\delta(\cdot)$ is the Dirac delta function.

Applying the E-step, which is to take the expected value of the log-likelihood in (\ref{eq:s2_loglik}), with respect to the complete conditional distribution of each $Z_i$, we obtain
\begin{align}
    \label{eq:s2_estep}
\begin{split}
    Q(\cdot | \bm{y}_1, \dots, \bm{y}_n) = & \sum_{i=1}^n \sum_{j=1}^k \gamma_{ij} \left(\ln \eta_{ij} + \ln N_m(\bm{y}_{it} ; \bm{F}_{j} \bm{\theta}_{jt}, \bm{V}_{j})\right) + \\ & \sum_{j=1}^n \sum_{t=2}^T \ln N_{p_j}(\bm{\theta}_{jt}; \bm{G} \bm{\theta}_{j,t-1}, \bm{W}_{jt})
\end{split}
\end{align}
where $\gamma_{ij}$ equals exactly the equation in (\ref{eq:s2_dummybayes}), but evaluated with the parameter values computed in the previous iteration of the algorithm. Note that by having the same form, it presents the same numerical problems as equation (\ref{eq:s2_dummybayes}); thus, the same care must be taken.

Then, the M-step revolves around maximizing the function in (\ref{eq:s2_estep}). Given that function (\ref{eq:s2_estep}) presents linearly independent terms, we can break down the optimization step into seperate blocks.

Firstly, the maximization for each $\bm{\eta}_{i}$ is linearly independent of everything else, with an expression of the form
\begin{equation*}
    \sum_{j=1}^k \gamma_{ij} \ln \eta_{ij}
\end{equation*}
for which the optimum is, unsurprisingly, exactly at $\eta_{ij} = \gamma_{ij}$.

The optimization for the parameters of each cluster can also be performed independently from the parameters of all other clusters, each with an expression of the form
\begin{equation}
    \label{eq:s2_weighteddlmloglik}
    \sum_{i=1}^n \gamma_{ij} \ln N_m(\bm{y}_{it} ; \bm{F}_{j} \bm{\theta}_{jt}, \bm{V}_{j}) + \sum_{t=2}^T \ln N_{p_j}(\bm{\theta}_{jt}; \bm{G} \bm{\theta}_{j,t-1}, \bm{W}_{jt})
\end{equation}
which is the weighted log-likelihood of an usual Dynamic Linear Model.

This is the time note \emph{why} the use of discount factors is so important for this model. We essentially have that the log-likelihood for a DLM is a weighting of two types of errors:
\begin{enumerate}
    \item The \emph{observational error}, given by the first term of (\ref{eq:s2_weighteddlmloglik}) and weighted by its precision terms $\bm{V}_j^{-1}$, and;
    \item The \emph{evolutional error}, given by the second term of (\ref{eq:s2_weighteddlmloglik}), weighted by its precision terms $\bm{W}_j^{-1}$.
\end{enumerate}

As far as maximum likelihood (or maximum a posteriori) estimates are concerned, assuming a constant $\bm{W}_j$ which is not a function of anything else leads to a divergent optimization problem. Essentially, the optimum will be found when minimizing one of the errors while throwing the respective variance to near zero while setting the other variance to a very high value, tending to infinity. More concretely, either the $\bm{\theta}_{l}$ will overfit the data with an estimation of $\bm{V}_j$ going to zero, or the $\bm{\theta}_{l}$ are going to overfit the evolutional equation with an estimation of $\bm{W}_j$ going to zero.

By using discount factors, each $\bm{W}_{jt}$ becomes a function of the values of $\bm{V}_j$ in such a way to guarantee a certain level of smoothness to the evolution of $\bm{\theta}_j$, with the level of smoothness being dictated by the discount factor itself. Because of the imposition of this relation, the optimum no longer lies in the previously detailed edge cases, and useful estimations can be obtained.

We still need, however, to be able to optimize the weighted DLM. For this, we use coordinate descent \citep{wright2015coordinate} for the optimization of each individual DLM. This allows us to iteratively optimize $\bm{\theta}_j$ given $\bm{V}_j$ and $\bm{V}_j$ given $\bm{\theta}_j$. This is useful because we are able to obtain analytical solutions for each of these steps, and consequently this will run much more efficiently than a usual gradient descent approach, for example.

The optimization of $\bm{\theta}_j$ given $\bm{V}_j$ can be done by tiling all of the observations as mentioned before, considering observations $\bm{y}_t$ which are a vertical concatenation of all $\bm{y}_{it}$ and considering an observational matrix which is a vertical tiling of $n$ replicates of the original $\bm{F}_j$. The difference is that here the observational error covariance matrix is given by $\bm{I}^{(j)} \otimes \bm{V}_j$ where, instead of using the identity matrix we use, $\bm{I}^{(j)} = diag(\gamma_{1j}^{-1}, \dots, \gamma_{nj}^{-1})$. This form for the observational covariance matrix takes into account the weights of the observations by inflating the variances for entries with low weights. Then, instead of running the standard Forward Filtering Backwards Sampling procude we perform Forward Filtering and in the backwards steps, instead of obtaining samples we obtain the modes.

When the weight of some observations are exceedingly small for a certain cluster, the matrix $\bm{I}^{(j)}$ becomes numerically unstable. To avoid this, we recommend setting a small value $\epsilon_{w}$ and discarding all observations with a weight lower than for that estimation step. This avoids numerical instability and, since the removed observations had small weights, the resulting value should not be significantly affected regardless of their inclusion.

The optimization of $\bm{V}_j$ given $\bm{\theta}_j$ reduces itself to a classic weighted estimator of the variance of normal observations.


In conclusion, the E-step is direct and the M-step is mostly analytical, with a very efficient subprocedure to optimize $\bm{\theta}_j$ and $\bm{V}_j$ for each cluster. All in all, this is a very quick and theoretically robust way to obtain point estimates for the model parameters and states.

\subsection{Algorithm Initialization}

Both the Expectation Maximization and Gibbs sampling algorithms are iterative and require us to initialize them with starting values for the parameters. Although both methods have proven theoretical properties indicating that the initial value is of small importance\footnote{True for the EM algorithm only for unimodal likelihoods since it's only able to arrive at local maxima.}, smart initialization can avoid computational problems and speed up convergence.

Here, we propose an algorithm combining the clever idea of~\cite{smyth1997clustering} of picking individual time-series to initialize each cluster, but we both reduce considerably the computational cost of doing so and introduce some robustness to the method by using the \textit{k-means++} algorithm \citep{arthur2007k} for the selection of the candidates.

The \emph{k-means++} algorithm is traditionally used to initialize parameters which lie in the same space as the observations, which is not our case since for an $m$-dimensional observation, we have an arbitrary $p_j$-dimension space for each of the $j = 1, \dots, k$ clusters.

It works as follows: at first, an observation is randomly (and uniformly) chosen to be the first centroid. Then, a new centroid is chosen randomly with probabilities proportional to the distance between each candidate and the existing centroids. This is done until $k$ centroids are found.

A common first impression from this algorithm is that it will have a tendency to place the initial centroids together with outlier observations. In practice, outliers will have the highest distances to most other points, and therefore are indeed the most likely individual \emph{points} to be chosen, however \emph{regions} with a lot of points that are moderately distant from the existing centroids will present higher probabilities of being picked. This way, unless the specified value of $k$ is too high, it's always \textit{probable} that the new centroid will be chosen within a different cluster. For detailed properties and discussion regarding this method, refer to the original paper.

An important difference between our model and the original context in which the algorithm was proposed is that, unlike in the \emph{k-means} case where the ordering of the clusters is always irrelevant, in our model this \emph{may} be relevant whenever the specifications of the pair $(\bm{F}_j, \bm{G}_j)$ differ from each other.

Consider, for example, three well-defined clusters: two where series have a random walk behaviour and one where series have a linear trend behaviour. It would then be desirable to have one of the observed series with a linear trend to initialize the linear growth model, and two of the observed series with the random walk behaviour to initialize the random walk models.

The proposal from~\cite{smyth1997clustering} tackles exatcly this problem. It revolves around estimating all $k$ models for all $n$ observations, and picking the highest likelihood observation to initialize each of the $k$ clusters.

Our combined approach is as follows: we separate the $k$ different DLM models into $l \leq k$ different sets, where DLMs from each set present the exact same pair $(\bm{F}_j, \bm{G}_j)$. In the example above there would be two sets for the three clusters: the random walk set and the linear growth set.

Then, we perform the \emph{k-means++} algorithm for picking $k$ obsevations to be the initial representatives for each of the $k$ clusters. Then, we estimate only $l$ different models for each of the $k$ observations, first reducing the number of model estimates from $k \times n$ to $k \times k$, and further removing the redundancy of estimating equivalent models, i.e. models with the same specification of $(\bm{F}_j, \bm{G}_j)$, multiple times, reducing the number of model estimates to $l \times k$.

The estimations of single DLM observations are done following the computationally efficient gradient descent procedure already detailed previously, and given the initialized values of each cluster, the initialization of each $\eta_{ij}$ can be performed by applying equation (\ref{eq:s2_computational_formula}) without the $\eta_{ij}$ terms.

\subsection{The label switching problem}

Performing MCMC sampling while working with mixture models can lead to various problems regarding identifiability of the ordering of the clusters. See~\cite{diebolt1994estimation} and~\cite{fruhwirth2001markov} for detailed explanations.

For our model, as already mentioned, whenever DLMs have the same pair $(\bm{F}_j, \bm{G}_j)$ there is no observational information regarding an ideal order. Thus, a simple ordering restriction like the one detailed by~\cite{fruhwirth2001markov} needs to be imposed.

\begin{figure}[H]
    \centering

    \includegraphics[width=0.55\linewidth]{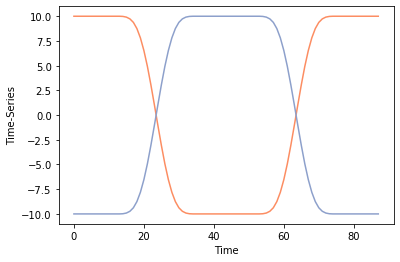}

    \caption{An example of two time-series with the same overall mean but wildly different shapes.}\label{fig:s2_restriction_example}
\end{figure}

However time-series may have, for example, the exact same overall means but different shapes; consider, for example, the series in Figure~\ref{fig:s2_restriction_example}. Assum they're the means for two clusters in a bigger modelling problem. If we tried to specify an ordering for the clusters through their overall mean, these two different behaviours would not differ from each other.

A practical solution that proved successfull is imposing the ordering restriction using as reference a single time instant. For the example pictured in Figure~\ref{fig:s2_restriction_example}, one could simply impose an ordering at, say, $t = 0$. Although theoretically one could always construct examples where there is no single time instant $t$ in which all $k$ clusters present a different level, this has proven to be a very successful strategy in practice.

\section{Dynamic Mixtures of Dynamic Linear Models}\label{sec:dynamic_mixture}

When the time window of observations is too long, it may be inappropriate to assume that an observation remains in the same cluster throughout all of it. An alternative is to allow observations to shift cluster memberships through time.

As far as the observational model is concerned the modifications are simple; we simply add a time index to the cluster membership parameters equation and equation (\ref{eq:s2_obs_weighted}) becomes
\begin{equation*}
    f(\bm{y}_{i} | \bm{\eta}_i, \bm{\theta}_1, \dots, \bm{\theta}_k, \bm{V}_1, \dots, \bm{V}_k) = \sum_{j=1}^k \prod_{t=1}^T \eta_{itj} N_m(\bm{y}_{it} ; \bm{F}_j \bm{\theta}_{jt}, \bm{V}_j)\text{,}
\end{equation*}
with the dummy variables also time-indexed so that $\bm{Z}_i = (Z_{i1}, \dots, Z_{iT})$, where, analogously, $\Pr(Z_{it} = j | \bm{\eta}_{it}) = \eta_{itj}$. Thus, equation (\ref{eq:s2_obs_dummy}) becomes
\begin{equation}
    \label{s3:independent_likelihood}
    f(\bm{y}_{i} | \bm{Z}_i, \bm{\theta}_1, \dots, \bm{\theta}_k, \bm{V}_1, \dots, \bm{V}_k) = \prod_{t=1}^T N_m(\bm{y}_{it} ; \bm{F}_{Z_{it}} \bm{\theta}_{Z_{it} t}, \bm{V}_{Z_{it}})\text{.}
\end{equation}

This, in of itself, is a completely valid model for which estimation is trivially done with a small extension on the methodology specified in the previous section. As a matter of fact, there exists a specific pair of $(\bm{F}_j, \bm{G}_j)$ such that, if applied to every cluster, would result in a model equivalent to that of~\cite{maruotti2016time}. This, this proposal is more general by allowing each cluster to possibly have a different DLM.

However, if no time-dependence is included between each $\bm{\eta}_{it}$ and $\bm{\eta}_{i,t-1}$, we will essentially have an independent clusterization at each time-point, which can lead to undesired effects such as having observations wildly switching cluster memberships because of single outlier observations\footnote{Again, an example of this will be shown in the applications}.

To avoid such problems we include temporal dependence between the weights by assuming that a Dirichlet Evolutional Model, as proposed by~\cite{fonseca2017dynamic}, determines the evolutional behavior of $\bm{\eta}_{it}$.

Noe that a consequence of $P(Z_{it} | \bm{\eta}_{it}) = \eta_{itj}$ is that $e_{Z_{it}} | \bm{\eta}_{it} \sim Multinomial(1, \bm{\eta_{it}})$. Considering these vectors as the latent observations for the Dirichlet evolution, we have from~\cite{fonseca2017dynamic} the following results.

\textbf{Definition 1 (Evolutional Dirichlet Process)}. Let $\bm{\psi}_{it} = (\psi_{it1}, \dots, \psi_{itk})'$ where the $\psi_{it1}, \dots, \psi_{itk}$ are independent with $\psi_{itj} \sim Beta(\delta_i c_{i,t-1,j}, (1-\delta_i) c_{i,t-1,j})$, where $\delta_i \in (0,1]$ is a discount factor and $c_{i,t-1,j} > 0$, $j = 1, \dots, k$. Let $S_{it} = \bm{\psi}'_t \bm{\eta}_{i,t-1}$ and let $\odot$ denote the entrywise product. Then, the Evolutional Dirichlet Process for $\bm{\eta}_{it}$ is defined by 
\begin{equation}
    \label{eq:edp_definition}
    \bm{\eta}_{it} = \frac{1}{S_{it}} \bm{\psi}_{it} \odot \bm{\eta}_{i,t-1}\text{.}
\end{equation}

The value of $\delta_i$ determines how volatile the cluster membership of the $i$-th observation is through time. Specifically, if $\delta_i \rightarrow 0, \forall i$ it would be the same as not considering temporal dependence, and if $\delta_i \rightarrow 1, \forall i$ it would be same the same as considering the static mixture from the previous section. Interest obviously lies on values of $\delta_i$ within the unit interval.

Let $\mathcal{D}_{i0}$ be the prior information about $\bm{\eta}_{i0}$ and $\mathcal{D}_{it} = \{\mathcal{D}_{i,t-1}, e_{Z_{it}}\}$ for $t > 0$.

\textbf{Theorem 1 (EDP Forward Filter).} Assume a prior distribution for the process of the form $\bm{\eta}_{i0} | \mathcal{D}_{i0} \sim Dirichlet(\bm{c}_{i0})$, consider $e_{Z_{it}} | \bm{\eta}_{it} \sim Multinomial(1, \bm{\eta}_{it})$ and the EDP given by (\ref{eq:edp_definition}). Then, for $t = 1, \dots, T$, given $\bm{\eta}_{i,t-1}|\delta_i, \mathcal{D}_{i,t-1} \sim Dirichlet(\bm{c}_{i,t-1})$, we have a prior distribution $\bm{\eta}_{it}|\delta_i, \mathcal{D}_{i,t-1} \sim Dirichlet(\delta_i \bm{c}_{i,t-1})$ and a posterior distribution $\bm{\eta}_{it}|\delta_i, \mathcal{D}_{it} \sim Dirichlet(\bm{c}_{it})$ where $\bm{c}_{it} = \delta \bm{c}_{i,t-1} + e_{Z_{it}}$.

It follows from Theorem 1 that $E[\bm{\eta}_{it} | \delta_i, \mathcal{D}_{i,t-1}] = E[\bm{\eta}_{i,t-1} | \delta_i, \mathcal{D}_{i,t-1}]$. Hence, it's important to note that the EDP implies a locally constant model.

\textbf{Theorem 2. (EDP Backwards Sampler)} To sample from $f(\bm{\eta}_{i1}, \dots, \bm{\eta}_{iT} | \delta_i, \mathcal{D}_{iT})$ we can first recursively compute the filtering distributions $\bm{\eta}_{it} | \delta_i, \mathcal{D}_{it} \sim Dirichlet(\bm{c}_{it})$ as given in Theorem 1 and sample from $\bm{\eta}_{iT} | \delta_i, \mathcal{D}_{iT} \sim Dirichlet(\bm{c}_{iT})$ directly. Then, for $t$ from $T$ to 2, recursively sample $\bm{\eta}_{i,t-1}$ given $\{\delta_i, \mathcal{D}_{iT}, \bm{\eta}_{it}, S_{it}\}$ which can be done as follows: sample $S_{it}$ given $\{\mathcal{D}_{iT}, \bm{\eta}_{it}, \delta_i\}$ which follows a $Beta(\delta_i \bm{1}'_k \bm{c}_{i,t-1}, (1-\delta_i) \bm{1}'_k \bm{c}_{i,t-1})$, sample $\bm{u}$ from $Dirichlet((1-\delta_i) \bm{c}_{i,t-1})$, and set $\bm{\eta}_{i,t-1} = S_{it} \bm{\eta}_{it} + (1 - S_{it}) \bm{u}$.

Note that even for high dimensional time-series, the parameter $eta_{i}$ is a $k$-variate series, with its trajectory over time is serving as a useful summary indicating the series' membership to each cluster thorough time, regardless of how high $m$ is.

\subsection{Gibbs sampling}

Firstly, sampling from the process $\bm{\eta}_i = \{\bm{\eta}_{i1}, \dots, \bm{\eta}_{iT}\}$ for each $i = 1, \dots, n$, given all the other parameters, particularly the dummy observations $Z_{it}$, can easily be performed through Forward Filtering Backwards Sampling as given by Theorems 1 and 2, respectively.

Sampling from the dummy variables $Z_{it}$ is done similarly to the way it was done in the previous section, except that now the weights are computed for only one time-instant at a time, removing the numerical problems regarding the large product of small values. Specifically, Equation (\ref{eq:s2_dummybayes}) becomes
\begin{equation*}
    P(Z_{it} = j | \bm{\eta}_{it}, \bm{\theta}_1, \dots, \bm{\theta}_k, \bm{V}_1, \dots, \bm{V}_k, \bm{y}_i) = \frac{\eta_{itj} N_m(\bm{y}_{it} ; \bm{F}_{j} \bm{\theta}_{jt}, \bm{V}_{j})}{\sum_{l=1}^k \eta_{itl} N_m(\bm{y}_{it} ; \bm{F}_{l} \bm{\theta}_{lt}, \bm{V}_{l})} \text{.}
\end{equation*}

Sampling from each of the DLM's specific parameters is also done in a very similar way. The only difference is that instead of each cluster having a fixed number of replications $n_j$, now it has a different number of replicates $n_{jt}$ at each time point. This is trivially solved by performing the tiling techniques for $\bm{F}$ and $\bm{V}$ at each time instant.

As for the $\delta_i$ parameter, which determine how volatile the evolution of the weights is for a single observational unit $i$, and therefore control how much an observational unit may switch clusters, can either be modelled or simply initialized.

As far as modelling is concerned, we can find that the marginal likelihood of $\delta_i$ given the dummy variables $Z_{i,1:T}$, is given by
\begin{equation*} \label{eq:delta_marginal}
    p(Z_{i,1:T} | D_{i,0}, \delta_i) = \prod_{t=1}^T \frac{\delta_i^{t-1} c_{i,0,Z_{it}} + \sum_{l=0}^{t-2} \delta_i^l \delta(Z_{i,t-1-l} - Z_{i,t})}{\delta_i^{t-1} \sum_{j=1}^k c_{i,0,j} + \frac{1 - \delta_{i}^{t-1}}{1-\delta_{i}}}\text{.}
\end{equation*}
Since this is a function with a domain within the unit interval, a Sampling Importance Resampling approach can be used to generate from this conditional distribution, using an uniform distribution, for example, as a proposal.

\subsection{Point Estimates}

The Expectation-Maximization algorithm has the property of effectively removing the dummy variables from the optimization through the Expectation step, substituting them with their expected value. Although useful for the static model, this property is not valuable in the dynamical case because the big advantage of using the Evolutional Dirichlet Process is the analytical Forward Filtering Backwards Sampling procedure, which relies on the presence of these latent observations.

Specifically, given the dummy variables, instead of Foward Filtering Backwards Sampling we can perform the aforementioned maximization procedure: performing usual Forward Filtering following Theorem 1 and, from Theorem 2, instead of sampling from the given distributions, simply obtaining the modes in a backwards fashion.

Inclusion of these non-continuous variables directly into the maximization procedure, however, is also unfeasiable. Optimization within continuous spaces are much simpler and efficient.

To solve this, we use the Stochastic Expectation-Maximization algorithm \citep{nielsen2000stochastic}. Essentially, in the step of maximizing the weights, we marginalize the dummy variables through a Monte Carlo approach using the same conditional distribution we would normally use to perform the expectation step in a traditional EM application. By doing this we guarantee the existence of the discrete values operationally when performing FFBS, but, at the same time, remove the necessity to directly include their estimation in the procedure.

As far as cluster specific estimation, we perform the same weighted DLM optimization, only now at each time-instant there is a different observational error covariance matrix since the membership weights change through time.

The iterative algorithm implemented is as follows, described as in an arbitrary iteration step:

\begin{enumerate}
    \baselineskip18pt
    \item {
        For each $i$:
        \vspace{-10pt}
        \begin{enumerate}
            \item Simulate $M$ different $\{\bm{Z}_{i1}, \dots, \bm{Z}_{it}\}$ from its conditional distribution evaluated with the last iteration's parameters.

            \item Take the mean of the $M$ different results as the Monte Carlo estimator for the result of maximizing the weights marginally on the dummy variables.
        \end{enumerate}
    }

    \item {For each $j$, perform direct maximization of the weighted DLM by coordinate descent.}
\end{enumerate}

If one wishes to perform joint estimation of $\delta$, then after step 1.(a), the function presented in (\ref{eq:delta_marginal}) can be be easily be optimized, for example, through a grid method. 

\subsection{Initialization}

Initialization for these methods are done using the same philosophy from the previous section. First, we initialize the DLM states and precisions following the exact same approach. Then, to initialize the weights given the initialized cluster parameters, we set
\begin{equation*}
    \eta_{itj} = \frac{N_m(\bm{y}_{it} ; \bm{F}_{j} \bm{\theta}_{jt}, \bm{V}_{j})}{\sum_{l=1}^k N_m(\bm{y}_{it} ; \bm{F}_{l} \bm{\theta}_{lt}, \bm{V}_{l})} \text{.}
\end{equation*}

Although Bayesian estimation of the Evolutional Dirichlet Evolution discount factors $\delta_{i}$ can be performed as mentioned in~\cite{fonseca2017dynamic}, we are using an intuitive algorithm to initialize them.

The basic idea involves quickly estimating the dynamic mixture model, considering no time-dependencies between the weights, which is computationally very fast to do. With the results, it is easy to check through the computed weights if a time-series is likely to change clusters. If it is, $\delta_i$ is set to a moderate value and if the time-series, even in the independent weights model, generally stayed in the same cluster, $\delta_i$ is set to a value close to 1.

\section{Applications}

Two applications are presented to illustrate how the model performs in a real dataset with univariate and multivariate time-series. Firstly, for the univariate example, World Bank indicators regarding the usage of renewable energy by EU countries was considered. For the second example, the popular Gapminder dataset was used, providing life-expectancy and GDP per capta for various countries, from which European and African ones were selected to better illustrate dynamic cluster memberships of certain nations.

\subsection{Renewable energy consumption in the EU}

One of the big topics of discussion of the European Comission (EC) relates to usage of Renewable energy from the member states. This is an important topic to them not only because of the gain in sustainability and the technological advances that result from investments in renewable energy sources, but it also presents the strategic and economic benefit of relying less on the imports of natural gas and oil. In 2017, 55.1\% of the gross available energy resources in the EU came from imported sources \citep{EnergyEC}.

As such, in 2004 the EU Directive on Electricity Production from Renewable Energy Sources (2001/77/EC), popularly known as the RES directive, came into effect, setting national targets for renewable energy production from individual member states, setting targets for 2010 and 2020.

In 2009, the RES directive was superseded by the ongoing Renewable Energy Directive (2009/28/EC), which sets more extensive targets, with the aim that in 2020, 20\% of total energy consumption within the EU will come from renewable resources. This directive obliged countries to present, by 2010, a National Renewable Energy Action Plan, explaining the country's road map to achieving its goal. In 2015, the European Comission published that they are on track to achieve the directive's goals \citep{EnergyOnTrack}.

\begin{figure}[H]
    \centering

    \includegraphics[width=\linewidth]{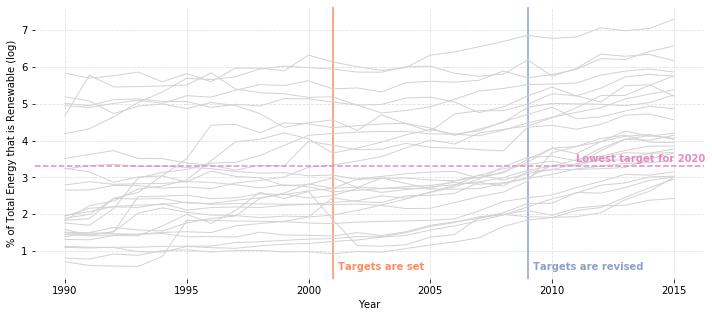}
    \caption{In 2001 renewabe targets were set and in 2009 they were revised. We can see that regardless of the country, shortly after 2005 renewable energy usage started to grow.}\label{fig:eu_data}
\end{figure}

The World Bank publishes indicator for various nations, including ones from the European Union, regarding renewable energy consumption as a percentage of the country's total final energy consumption \citep{WorldBankEU}. This data is available for all EU countries between 1990 and 2015, with an yearly resolution. Due to Malta's relative unimportance to the overall goals because of its size, and because until 2002 it had no renewable energy consumption, it has been excluded from the analysis. This data is shown in Figure~\ref{fig:eu_data}.

\begin{figure}[H]
    \centering

    \includegraphics[width=\linewidth]{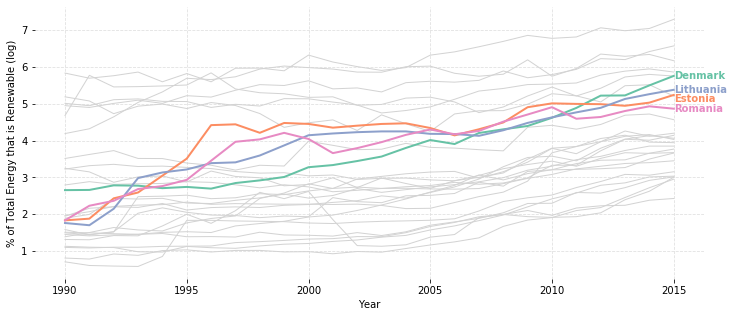}
    \caption{Estonia, Lithuania and Romania had growing rates of renewable energy usage since the start of the time-window at 1996, whereas Denmark started a more rapid growth at around 1999.}\label{fig:eu_growers}
\end{figure}

We can see that there is a large group of countries with very low rates of Renewable Energy Usage (REU), a smaller group with higher REU and a few countries that showed some growth during this time-period, moving from the lower group to the higher one between 1995 and 2010. These countries are highlighted in Figure~\ref{fig:eu_growers}.

Denmark, by 2015, was already the EU country with the fifth-highest REU. Its growth was mainly due to Denmark effecient use of wind energy production. According to Denmark's own Energinet, in 2017 81,4\% of its energy came from renewable resources, up from 8.7\% in 1997. In 2017, an astounding 50.2\% of its energy production came solely from wind \citep{Energinet}.

Lithuania's growth has been largely due to its extensive use of biomass resources \citep{Lithuania}, Estonia has relied more heavily on wind \citep{Estonia} and Romania has relied morsly hydro-energy (19\% of total energy in February, 2019) but with significant shares of wind and solar (9\% and 8\% of total, respectively, in February, 2019) \citep{Romania}.

\begin{figure}[H]
    \centering

    \includegraphics[width=\linewidth]{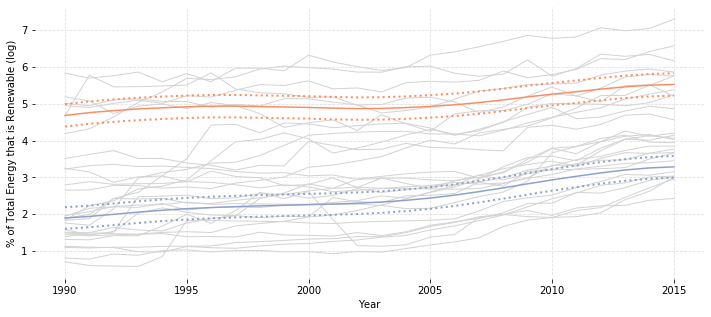}
    \caption{Modelled means for each of the clusters, with dashed lines indicating two standard deviations.}\label{fig:eu_means}
\end{figure}

We fitted the model presented in Section~\ref{sec:dynamic_mixture} to this data considering $k = 2$ clusters and assuming a random-walk behavior for each of them. The resulting means for each of the clusters are shown in Figure~\ref{fig:eu_means}.

\begin{figure}[H]
    \centering

    \includegraphics[width=\linewidth]{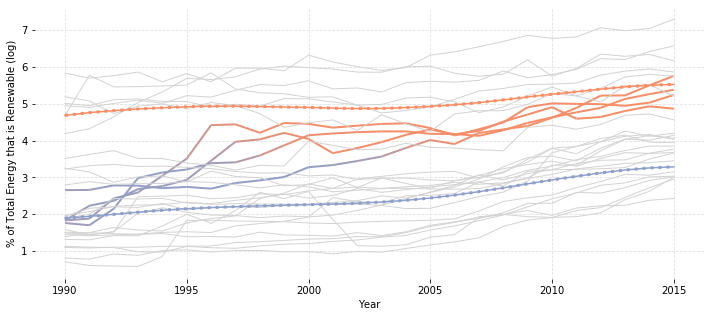}
    \caption{Classification for the countries with growth behavior.}\label{fig:eu_class_grow}
\end{figure}

We can see that the model successfully captures the general behavior of both clusters. Not only that, but we can see the series for the growing countries in Figure~\ref{fig:eu_class_grow}, highlighted with a color reflecting their membership weights at each time-instant.

From the model results, we have that Estonia, the earliest grower, had uncertain membership weights around 1995 and 1996, whereas Romania and Estonia switched cluster behaviour at about 1997 and 1998, respectively. Denmark, the latest grower, changed cluster behaviour between 2002 and 2004, soon after the RES directive. Note that the proposed model allows for identifying the change points in time for the series moving from one cluster to another.

\begin{figure}[H]
    \centering

    \includegraphics[width=\linewidth]{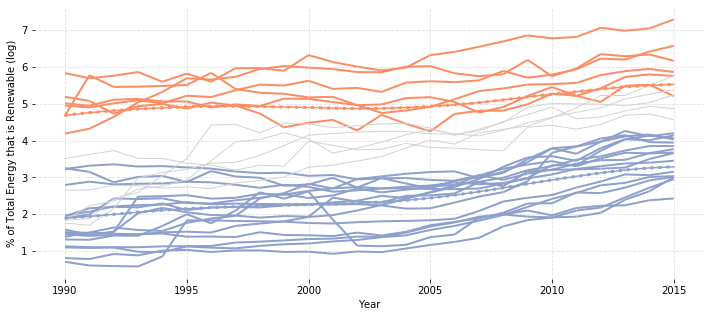}
    \caption{Classification for the countries with static behavior.}\label{fig:eu_class_other}
\end{figure}

From Figure~\ref{fig:eu_class_other} we can see that other countries were statically assigned to each of their clusters, with no significant changes in their memberships throughout the entire time-window, as expected.

\subsection{Life-expectancy and GDP per capta in Europe and Africa}

For this second analysis, we considered the famous Gapminder dataset informing GDP per capta and Life Expectancy from various nations which was used to create the popular presentation \emph{``How Does Income Relate to Life Expectancy?''} \citep{gapminder}.

\begin{figure}[H]
    \centering

    \includegraphics[width=\linewidth]{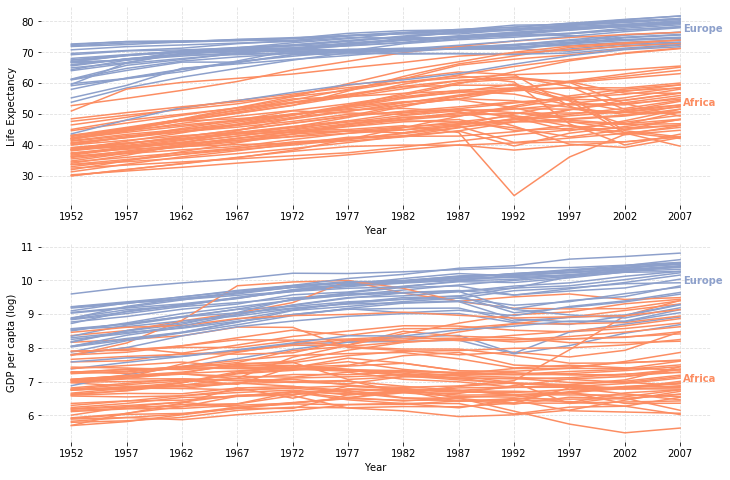}
    \caption{Gapminder data for Life Expectancy (above) and log-GDP per capta (below). European countries in blue and African countries in orange.}\label{fig:gapdata}
\end{figure}

To better illustrate and visualize cluster-changing behavior, we only selected nations classified as either African or European from the original 142 countries, resulting in a selection of 82 bivariate time-series from both continents. The dataset presents information for every five years between 1952 and 2007. We can visualize the data in Figure~\ref{fig:gapdata}.

We can see from Figure~\ref{fig:gapdata} that, in general, all countries showed significant improvements in life expectancy, and most countries had considerable increase in GDP per capta, though with a number of exceptions. We can further see that, unsurprisingly, European countries tend to have much higher Life Expectancy \textit{and} GDP per capta than African counterparts.

It's evident that there are two groups of behavior: the European group and the African group. However, paying close attention to the graph, we can see that there are exceptions to this rule.

The two orange curves in the Life Expectancy graph which are clearly among the European group refer to Réunion, which is an overseas department and region of France, and the Republic of Mauritius, which is another island nation near Réunion. In the same graph, the blue line which starts out among the African group refers to Turkey, which showed significant growth.

As far as GDP per capta is concerned, the group divisions are less clear. However we have three African curves which, specially between 1962 and 1987, stay closely related to the European group; these refer to South Africa, which is a very rich African nation, and Lybia and Gabon, which are nations that rely strongly on natural resource extraction, most of that being oil. European countries with low GDP are Albania, Bosnia and Herzegovina, and Turkey, which shows a similar behavior for GDP per capta as it does for Life Expectancy.

\begin{figure}[H]
    \centering

    \includegraphics[width=\linewidth]{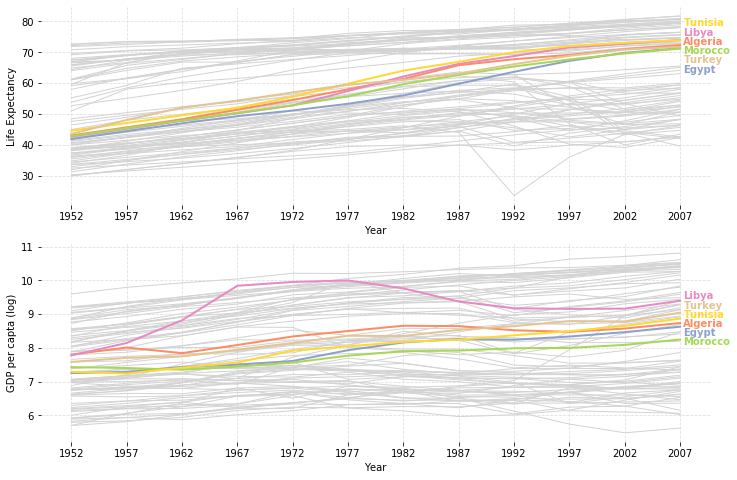}
    \caption{Gapminder data for Life Expectancy (above) and log-GDP per capta (below) for North African nations and Turkey.}\label{fig:natgrowth}
\end{figure}

Most interestingly, however, are the North African nations, along with Turkey, which showed considerable growth in Life Expectancy and very subtle growth of GDP per capta, as highlighted in Figure~\ref{fig:natgrowth}. Out of these countries, Morroco showed the least improvements in both Life Expectancy and GDP per capta.

\begin{figure}[H]
    \centering

    \includegraphics[width=\linewidth]{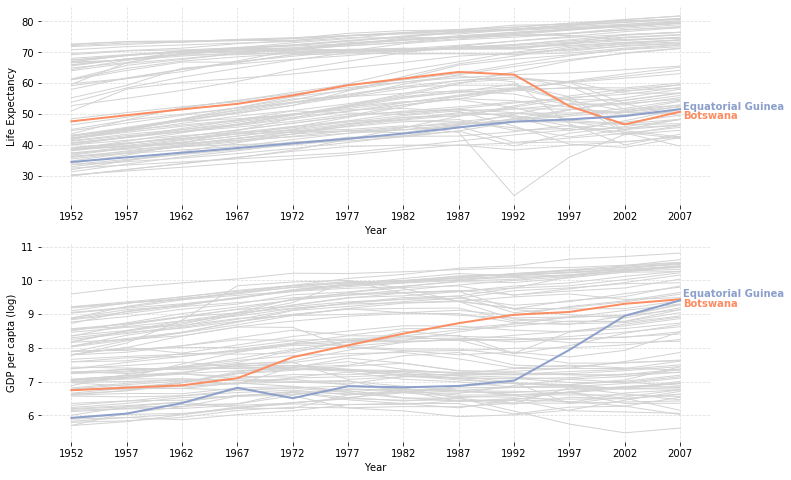}
    \caption{Gapminder data for Life Expectancy (above) and log-GDP per capta (below) for Botswana and Equatorial Guinea.}\label{fig:gdpgrowth}
\end{figure}

Two countries showed strong improvement in GDP per capta, as highlighted in Figure~\ref{fig:gdpgrowth}: Botswana, which presented steady growth since around 1970, and more evidently Equatorial Guinea, which had a significant jump from 1992 to 1997 due to the major economic development resulting from the discovery of large oil reserves in 1996 \citep{opec}.

We can see that Botswana after around 1990, along with various other sub-saharan nations, presented a very large drop in Life Expectancy, which only started to recover around 2000. This decade was marked by a large and deadly HIV epidemic in Africa, from which some countries have not been able to fully recover to this day, as highlighted by their Life Expectancies.

In a consumer-grade notebook, pontual estimation for this dataset with $n = 82$ different $m = 2$ dimesional time-series accross $T = 12$ time-instants took about 2 minutes, whereas the MCMC approach took about ten times as long. Assuming the existance of $k = 2$ different clusters we obtained means which are shown below in Figure~\ref{fig:gapmeans}.

\begin{figure}[H]
    \centering

    \includegraphics[width=\linewidth]{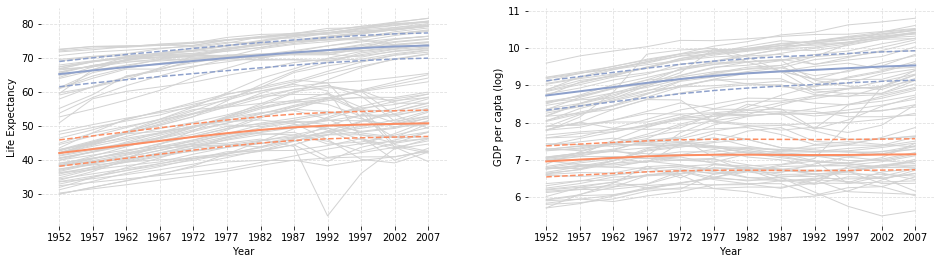}
    \caption{Estimated means for each of the clusters with an interval of two standard deviations.}\label{fig:gapmeans}
\end{figure}

\begin{figure}[H]
    \centering

    \includegraphics[width=\linewidth]{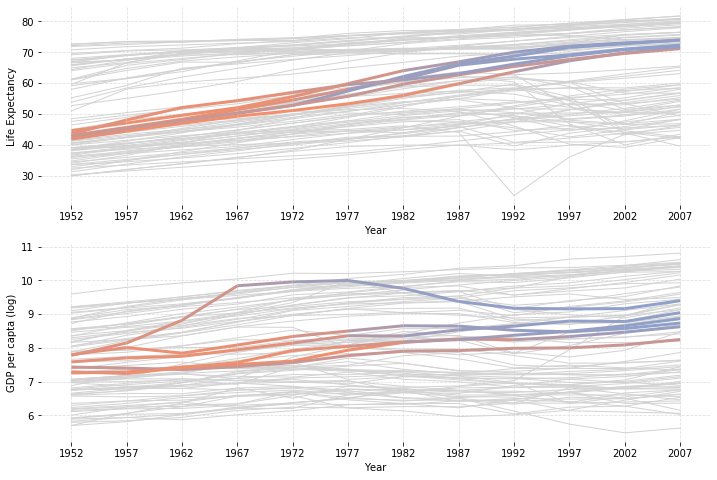}
    \caption{Cluster classifications for North African countries and Turkey.}\label{fig:natclass}
\end{figure}

Due to this being a bivariate application, classifications are considerably more interesting since the model has to weight changes in behavior for two different variables. As it happens, only the North African nations and Turkey actually presented a change in their classification, with the exception of Morrocco, as shown below in Figure~\ref{fig:natclass}, which had its cluster membership to the European group estimated at 45\% in 2017.

We can see that although Lybia had a strong GDP per capta, in 1967 it only had an estimated 45\% membership to the European group due to the low Life Expectancy. However in 1977 it already had an 88\% membership, being classified as en European-level country before the other North African nations due to the help of its GDP per capta.

Although Equitorial Guinea and Botswana had strong GDP per capta by 2007, their very low Life Expectancies kept them strongly within the African group, with estimated memberships nearing 100\%.

\begin{figure}[H]
    \centering

    \includegraphics[width=\linewidth]{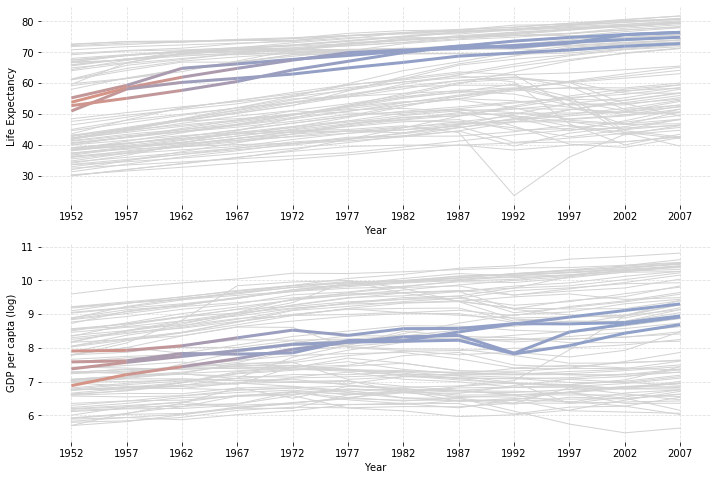}
    \caption{Cluster classifications for North African countries and Turkey.}\label{fig:uncertainbeginning}
\end{figure}

Of note, a few countries that were mostly classified as part of the European group had uncertain classifications in the first few data-points. These are: Albania, Bosnia and Herzegovina, Mauritius and Réunion. They are highlighted in Figure~\ref{fig:uncertainbeginning}.

Their memberships in the 1952 are estimated around 50\%, meaning that for these time-points, as far as the model is concerned, they could be classified as a member of either of the groups, but these rates quickly go up and they end up being generally classified as European countries.

Every other nation not specifically mentioned received a static classification in the expected cluster as expected.

\section{Concluding remarks}

The Static Mixture of Dynamic Linear Models, although not the main result from this research, presents itself as an extremely efficient and flexible way to statically cluster time-series data, in the form of the EM-algorithm, whenever dynamix mixtures are found to be unecessary.

The Dynamic Mixture of Dynamic Linear Models has proven to be able to capture change of behavior from time-series while also correctly classifying time-series which clearly belong to the same cluster throughout the entire time-window. In particular, it is a demonstrable improvement over existing methods in that it avoids missclassification of outliers by employing an Evolutional Dirichlet Process to capture time-dependence of the cluster membership parameters.

This is also useful for dealing with a high number of high-dimensional time-series which can be segmented into a few different clusters. The general behavior of these series get summarised into the $k$-dimensional membership parameters $\eta$, which can be used for monitoring the series behaviour along with the $k$ different means, instead of going through the effort of analysing the actual $m$-variate time series when $m$ is very large.

A natural next step for development of this model is to attempt to incorporate not only a method for deciding on number of clusters, $k$, automatically, but also to allow $k$ to evolve over time, to encompass, for example, a set of observations which all present the same behaviour until at a time-instant when they separate themselves into a few different clusters.


\baselineskip11pt
\bibliographystyle{apalike}


\newpage

\appendix

\section{Application to Artificial Data}\label{sup:app_to_art_dat}

\begin{figure}[H]
    \centering

    \includegraphics[width=0.5\linewidth]{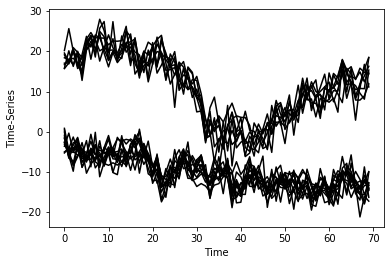}
    \caption{An artificial dataset for static clusterization of time-series.}\label{fig:s4_static_dataset}
\end{figure}

Consider the dataset presented in Figure~\ref{fig:s4_static_dataset}. In it, there are a total of 20 time-series which can be clearly differentiated into two different behaviours.

\begin{figure}[H]
    \centering

    \includegraphics[width=0.95\linewidth]{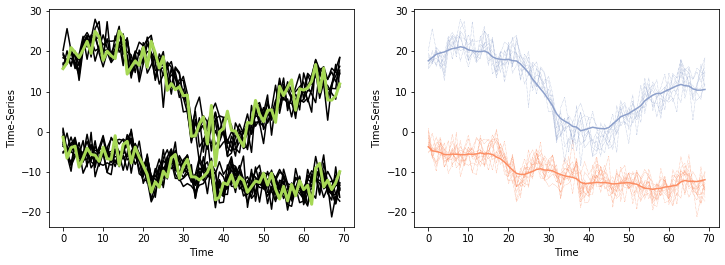}
    \caption{The two time-series selected to initialize the two different clusters are shown in green (left) and the result of the initialization procedure, where the color of a time-series represents its classification (right).}\label{fig:s4_initialization}
\end{figure}

The initialization algorithm described in Subsection 2.3 chose initial time-series to represent each of the clusters as indicated in Figure~\ref{fig:s4_initialization}. The initialization then led to the states and classifications shown also in Figure~\ref{fig:s4_initialization}. We can see that the initialization algorithm performs wonderfully, and in effect the estimation procedure will serve only to smooth out the estimates for the states and possibly correct small missclassifications.

\begin{figure}[H]
    \centering

    \includegraphics[width=0.95\linewidth]{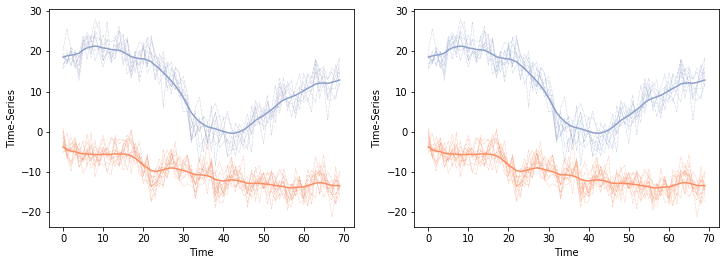}
    \caption{The result of estimation by the Gibbs sampler (left) and EM algorithm (right). The color of the time-series represent its classification.}\label{fig:s4_static_result}
\end{figure}

Running in a regular consumer-grade notebook, the pontual estimation took 5 seconds to run whereas the Gibbs sampler took 9 minutes and 38 seconds. The results from both are shown in Figure~\ref{fig:s4_static_result}, where it can be seen that the estimates from both methods are essentially the same, as one would expect.

\begin{figure}[H]
    \centering

    \includegraphics[width=0.55\linewidth]{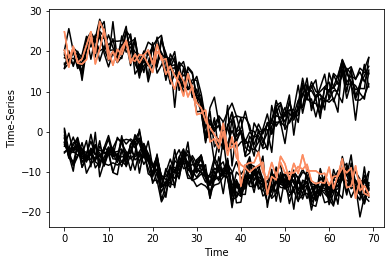}
    \caption{The artificial dataset for dynamic clusterization of time-series. The two time-series in orange were added to the dataset to represent the membership changing behaviour.}\label{fig:s4_dynamic_dataset}
\end{figure}

Consider now the dataset shown in Figure~\ref{fig:s4_dynamic_dataset}, where two time-series were added in such a way that they follow the behavior from the first cluster up until the point where the two clusters nearly meet, at which point they change their membership to the second cluster.

The proposed algorithm for picking discount factors returned a value of 0.95 for all of the original observations, and a discount factor of 0.5 for the new membership-changing time-series.

\begin{figure}[H]
    \centering

    \includegraphics[width=\linewidth]{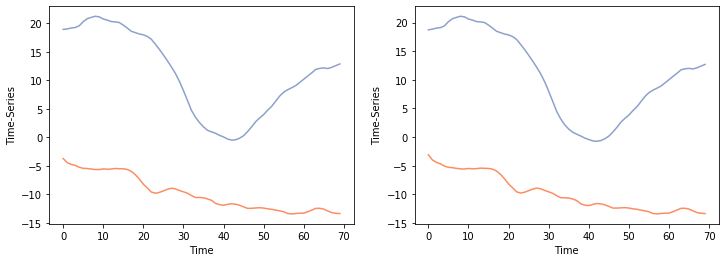}
    \caption{Estimation result of the Gibbs sampler (left) and pontual estimation algorithm (right) for the states.}\label{fig:s4_dynamic_theta}
\end{figure}

\begin{figure}[H]
    \centering

    \includegraphics[width=\linewidth]{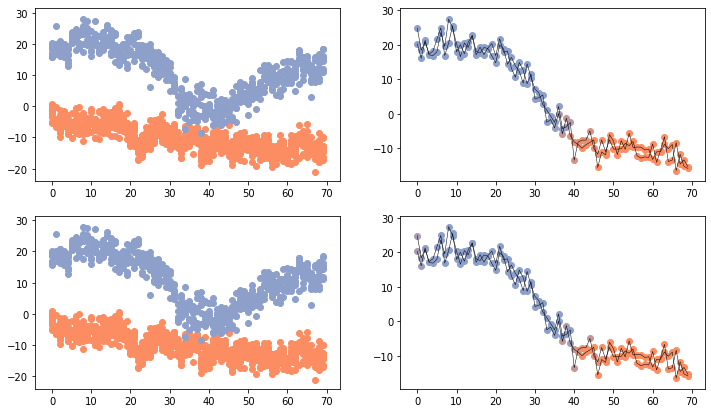}
    \caption{Classification results for the original 20 observations (left) and the two new observations (right) from the Gibbs algorithm (top) and the pontual estimation algorithm (bottom).}\label{fig:s4_dynamic_classification}
\end{figure}

The pontual estimation implementation took 1 minute and 18 seconds to run and the Gibbs sampler took 14 minutes and 23 seconds to run. In Figure~\ref{fig:s4_dynamic_theta} we can see the estimation result for the states from both algorithms. In Figure~\ref{fig:s4_dynamic_classification} we can see the classification results from both algorithms.

As expected, both algorithms give approximately the same results, the original 20 observations are still fully classified into one of the two clusters, and the two new additions have their cluster membership changing along with their behaviour.

\begin{figure}[H]
    \centering
    
    \includegraphics[width=\linewidth]{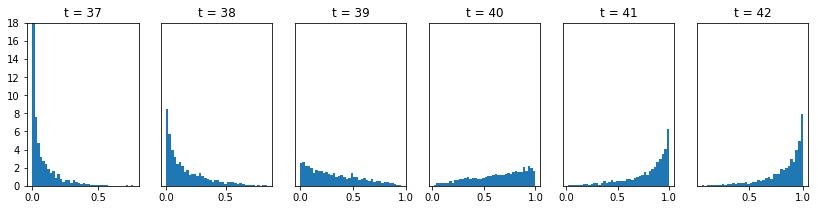}
    \caption{Posterior distribution for the $\eta$ parameter of one of the time-series that changed clusters, from $t = 37$ to $t = 42$.}\label{fig:s4_change_posterior}
\end{figure}

We can visualize the classification uncertainty by looking at the posterior distribution for the $\eta$ parameter for one of the cluster changing time-series around $t = 39$, when the cluster change actually took place. The distributions are highlighted in Figure~\ref{fig:s4_change_posterior}.

One important thing to note is the importance of including the Evolutional Dirichlet Process. The developed package also implements the case mentioned in Equation (\ref{s3:independent_likelihood}), where the cluster membership is independent through time.

\begin{figure}[H]
    \centering

    \includegraphics[width=\linewidth]{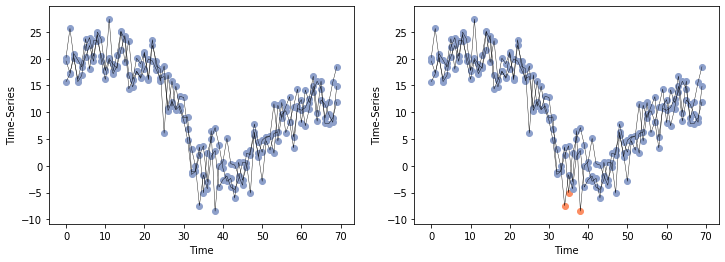}
    \caption{Higher-valued cluster time-series classification from the model with Evolutional Dirichlet Process for the weights (left) and indendent weights (right).}\label{fig:s4_missclassification}
\end{figure}

In Figure~\ref{fig:s4_missclassification} are examples of misclassifications resulting from the lack of dependence between time instants, where single outliers get incorrectly classified as members of the lower cluster because they don't have access to neighboring information about the time-series.

According to the Raftery diagnostic, the chains converged almost immedietaly, thanks to the initialization procedures, and the sampled values were only very lightly correlated.

\end{document}